\documentclass[letterpaper,twocolumn,prl,
aps,showpacs,superscriptaddress,floatfix]{revtex4-2}

\usepackage[latin1]{inputenc}
\usepackage{bbm}
\usepackage{bm}
\usepackage[usenames]{color}
\usepackage{multirow}
\usepackage{amssymb}
\usepackage{amsbsy}
\usepackage{mathtools}
\usepackage{amsmath}
\usepackage{stmaryrd}
\usepackage{graphicx}
\usepackage{epsfig}
\usepackage{placeins}
\usepackage{bbold}
\usepackage{braket}
\usepackage{blindtext}
\usepackage[colorlinks,linkcolor=blue,citecolor=blue,urlcolor=blue]{hyperref}

\usepackage{scalerel}
\usepackage{tikz}
\usetikzlibrary{calc}
\usetikzlibrary{patterns}
\usetikzlibrary{svg.path}
\definecolor{orcidlogocol}{HTML}{A6CE39}
\tikzset{
  orcidlogo/.pic={
    \fill[orcidlogocol] svg{M256,128c0,70.7-57.3,128-128,128C57.3,256,0,198.7,0,128C0,57.3,57.3,0,128,0C198.7,0,256,57.3,256,128z};
    \fill[white] svg{M86.3,186.2H70.9V79.1h15.4v48.4V186.2z}
                 svg{M108.9,79.1h41.6c39.6,0,57,28.3,57,53.6c0,27.5-21.5,53.6-56.8,53.6h-41.8V79.1z M124.3,172.4h24.5c34.9,0,42.9-26.5,42.9-39.7c0-21.5-13.7-39.7-43.7-39.7h-23.7V172.4z}
                 svg{M88.7,56.8c0,5.5-4.5,10.1-10.1,10.1c-5.6,0-10.1-4.6-10.1-10.1c0-5.6,4.5-10.1,10.1-10.1C84.2,46.7,88.7,51.3,88.7,56.8z};
  }
}

\newcommand\orcid[1]{\!%
  \href{https://orcid.org/#1}{%
    \mbox{%
      \scaleto{%
        \begin{tikzpicture}[yscale=-1,transform shape]
          \pic{orcidlogo};
        \end{tikzpicture}
      }{8pt}%
    }%
  }%
}

\makeatletter

\newcommand{\be}{\begin{equation}}
\newcommand{\ee}{\end{equation}}

\makeatletter

\begin{document}
\title{
Local integrability breaking and exponential localization of leading Lyapunov vectors
}
 \date{\today}
\author{Jiaozi Wang~\orcid{0000-0001-6308-1950}}
\affiliation{Department of Mathematics/Computer Science/Physics, University of Osnabr\"uck, D-49076 
Osnabr\"uck, Germany}
\author{Toma\v{z} Prosen~\orcid{0000-0001-9979-6253}}
\affiliation{Faculty of Mathematics and Physics, University of Ljubljana, Jadranska 19, SI-1000 Ljubljana, Slovenia}
\affiliation{Institute of Mathematics, Physics and Mechanics, Jadranska 19, SI-1000 Ljubljana, Slovenia}
\author{Giulio Casati~\orcid{0000-0002-8788-3090}}
\affiliation{Center for Nonlinear and Complex Systems, Dipartimento di Scienza e Alta Tecnologia,
	Universit\`a degli Studi dell'Insubria, via Valleggio 11, 22100 Como, Italy}

\begin{abstract}
We study integrability breaking and transport in a discrete space-time lattice with a local integrability breaking perturbation.  
We find a singular distribution of the Lyapunov spectrum where the majority of Lyapunov exponents vanish in the thermodynamic limit. The sub-extensive sequence of nonzero exponents, converging in the thermodynamic limit, correspond to Lyapunov vectors that are exponentially localized with localization lengths proportional to inverse Lyapunov exponents. Moreover, we investigate the transport behavior of the system by considering the spin-spin and current-current spatio-temporal correlation functions. Our results indicate that the overall transport behavior, similarly as in the purely integrable case, conforms to Kardar-Parisi-Zhang scaling in the thermodynamic limit and at vanishing magnetization. The same dynamical exponent $z=3/2$ governs the effect of local perturbation spreading in the bulk.

\end{abstract}
\maketitle
{\it Introduction.--} The widespread phenomenon of deterministic chaos has traditionally been investigated from the perspective of a few body dynamics. The famous Kolmogorov-Arnold-Moser~\cite{Kolmogorov,Arnold} theorem, the fundamental mathematical statement on stability of integrable classical Hamitonian dynamics to perturbations, assumes that the number of degrees of freedom $N$ is fixed. However, the behavior of integrability breaking thresholds on $N$ remained illusive (see Volume \cite{FPU} and Refs. therein), or highly 
non-optimally bounded (cf. Nekhoroshev theory~\cite{Niederman2009}), even after decades of research.

Probably the most precise quantitative measure of chaotic dynamics is the Lyapunov spectrum of characteristic exponents~\cite{Politi}
$\{\lambda_k\}_{k=1}^{2N}$, canonically ordered as $\lambda_1\ge \lambda_2\ge\cdots$.
While the maximal Lyapunov exponents $\lambda_1$ gives the characteristic time-scale $1/\lambda_1$ of exponential sensitivity to initial conditions (aka butterfly effect), and the sum of positive exponents $K_{\rm KS} = \sum_{k=1}^N \lambda_k$ is the Kolmogorov-Sinai entropy related to minimal information resource required to encode dynamical trajectory per unit of time~\cite{Gaspard1998}, there is much more information in the structure of the distribution $P(\lambda) = \frac{1}{N}\sum_k \delta(\lambda-\lambda_k)$, in particularly in its scaling in the thermodynamic limit $N\to\infty$. Recently, it has been suggested that many-body Hamiltonian dynamics with global integrability breaking can exhibit two distinct behaviors, characterized by short-range or long-range coupling in the underlying action network 
\cite{FlachPRL2019,FlachPRE2019,FlachPRE2021,PhysRevLett.128.134102-Flach,PhysRevE.108.L062301_Flach,PhysRevResearch.6.L012064_Flach}.

Complementing the Lyapunov spectrum, the Lyapunov vectors serve as another important measure of chaotic dynamics, contributing to the identification of both the real-space structure of collective modes and the regions with stronger or weaker instabilities \cite{kaneko1986-LV,HChate_1993-LV,egolf2000mechanisms-LV,PhysRevLett.92.254101-LV}.
Previous studies have focused on cases where integrability is broken by global perturbations, i.e., perturbations acting on all degrees of freedom.
It is found that the (covariant) Lyapunov vectors  associated with collective modes and conservation laws are delocalized, whereas the remaining vectors are typically localized \cite{Ginelli_2013-CLV13,Arkady_Pikovsky_1998-CLV98,PhysRevLett.99.130601-CLV07,PhysRevLett.103.154103-CLV09,PhysRevLett.111.064101-CLV13,PhysRevE.68.046203}.

A fundamentally different case of local integrability breaking, which is supported
on a single (or a few neighboring) spatially localized and
locally coupled degrees of freedom, has been somewhat overlooked to date. Nevertheless, one can imagine many interesting physical situations where integrability is broken locally, e.g. considering impurities in soliton models. Several important questions remain open, such as the structure of the Lyapunov spectrum and Lyapunov vectors, and whether the local perturbation changes transport behavior in systems with conserved charges.
 
To address these questions, we consider as a prime example a space-time discrete Landau-Lifshitz magnet ~\cite{krajnik2020kardar} with a local magnetic field (impurity). 
In the absence of impurity, the model is integrable, possesses a non-abelian (SO(3)) symmetry and has been suggested to exhibit Kardar-Parisi-Zhang (KPZ) scaling of 2-point dynamical correlations~\cite{MarkoPRL2019,krajnik2020kardar,Enej2020,Takeuchi2024}. 
It becomes nonintegrable after introducing a local magnetic field,
and we find a finite number, or a vanishing fraction of non-vanishing Lyapunov exponents with the corresponding Lyapunov vectors exponentially localized around the integrability breaking impurity. Moreover, the localization lengths of those Lyapunov vectors are given by the corresponding inverse Lyapunov exponents. Furthermore, we investigate the transport behavior by analyzing the spin-spin and current-current correlation functions. We show that local integrability breaking does not modify overall transport behavior at a length scale which moves away from integrability breaking impurity as $t^{1/z}$, with KPZ dynamical exponent $z=3/2$. This finding is qualitative different from the case of global integrability breaking, where KPZ scaling has been found to eventually break down even under symmetry-preserving perturbations \cite{PhysRevB.110.L180301-KPZ,Roderich}.

{\it The model of local integrability breaking.-- }
We consider rotationally (SO(3)) symmetric discrete space-time lattice dynamics introduced in Ref.~\cite{krajnik2020kardar}. 
Letting $\boldsymbol{S}_1, \boldsymbol{S}_2$ denote a pair of three-dimensional unit vectors, one defines a family of local symplectic maps $(\boldsymbol{S}_{1}^{\prime},\boldsymbol{S}_{2}^{\prime})=\varPhi_{\tau}(\boldsymbol{S}_{1},\boldsymbol{S}_{2})$, 
\begin{align}
\boldsymbol{S}_{1}^{\prime} & =\frac{1}{\sigma^{2}+\tau^{2}}(\sigma^{2}\boldsymbol{S}_{1}+\tau^{2}\boldsymbol{S}_{2}+\tau\boldsymbol{S}_{1}\times\boldsymbol{S}_{2}), \nonumber \\
\boldsymbol{S}_{2}^{\prime} & =\frac{1}{\sigma^{2}+\tau^{2}}(\sigma^{2}\boldsymbol{S}_{2}+\tau^{2}\boldsymbol{S}_{1}+\tau\boldsymbol{S}_{2}\times\boldsymbol{S}_{1}),
\end{align}
where $
\sigma^{2}:=\frac{1}{2}(1+\boldsymbol{S}_{1}\cdot\boldsymbol{S}_{2})$.
Considering a lattice of an even number $L\in 2\mathbb{N}$ of unit vectors~\footnote{
Each $\boldsymbol{S}_x$ takes values on $3D$ unit sphere and constitute two canonical degree of freedom, hence $N\equiv 2 L$.} $\boldsymbol{S}^t_x$,
we define a dynamical map $\varPsi_\tau$: 
\begin{gather}
(\boldsymbol{S}_{0}^{t+1},\boldsymbol{S}_{1}^{t+1},\ldots,\boldsymbol{S}_{L-1}^{t+1})=\varPsi_{\tau}(\boldsymbol{S}_{0}^{t},\boldsymbol{S}_{1}^{t},\ldots,\boldsymbol{S}_{L-1}^{t}),\nonumber \\
\varPsi_{\tau}=\varPsi_{\tau}^{\text{odd}}\circ\varPsi_{\tau}^{\text{even}}=\prod_{x=0}^{L/2-1}\varPhi_{\tau}^{2x+1}\circ\prod_{x=0}^{L/2-1}\varPhi_{\tau}^{2x}.\label{eq-oe-map}
\end{gather}
$\varPhi_{\tau}^{x}$ indicates a local map applied to sites $x,x +1$ and we use periodical boundary condition ${\boldsymbol{S}}_{L+x}^{t}={\boldsymbol{S}}_{x}^{t}$. It has been shown in Ref. \cite{krajnik2020kardar} that the dynamics generated by $\varPsi_\tau$ is integrable.

To break the integrability, we introduce a local perturbation, i.e. a local linear rotation along the third axis: $\boldsymbol{S}^{\prime}=R^{\phi}(\boldsymbol{S})=U(\phi)\boldsymbol{S}$, where
\begin{equation}\label{eq-Uphi}
U(\phi)=\left(\begin{array}{ccc}
\cos\phi & -\sin\phi & 0\\
\sin\phi & \cos\phi & 0\\
0 & 0 & 1
\end{array}\right). 
\end{equation}
The perturbed many-body map can thus be defined as
\begin{equation}
\varPsi_{\tau}^{\phi}=R_{0}^{\phi}\circ\prod_{x=0}^{L/2-1}\varPhi_{\tau}^{2x+1}\circ R_{0}^{\phi}\circ\prod_{k=0}^{L/2-1}\varPhi_{\tau}^{2x},
\end{equation}
where $R^\theta _0$ indicates the local rotation at $x = 0$.

In the main text, we concentrate on the parameters $\tau = 1$ and $\phi = \frac{\pi}{2}$, while additional parameter settings are explored in~\cite{SM}.
It is straightforward to see that the total magnetization in the z-direction 
$S\equiv\sum_{x=0}^{L-1}S_{x}$ is conserved, where $S_{x}=\boldsymbol{S}_{x}\cdot\boldsymbol{e}_{3}$.
To account for the potential dependency of dynamical properties on conserved magnetization $\mu = S/L$, we employ a canonical invariant ensemble of initial conditions
$\rho_\mu^{\rm tot}\circ\varPsi_{\tau}^\phi = 
\rho_\mu^{\rm tot}$,
\be\label{eq-gc}
\rho_{\mu}^{\text{tot}}(\boldsymbol{S}_{0},\cdots,\boldsymbol{S}_{L-1})=\prod_{x=0}^{L-1}\rho_{\mu}(\boldsymbol{S}_{x}),
\ee
where
$ \rho_{\mu}(\boldsymbol{S})=\frac{1}{4\pi}\frac{\kappa}{\sinh\kappa}e^{\kappa S}
$, $\coth\kappa-1/\kappa=\mu$.
\textcolor{black}{It should be noted that, for convenience in the following discussion, we relabel positions with $x (x\ge \frac{L}{2})$ by $x - L$, and use $x=-\frac{L}{2}+1,-\frac{L}{2}+2,\ldots,\frac{L}{2}$ as  coordinate range.}
\begin{figure}[t]
	\includegraphics[width=1.0\columnwidth]{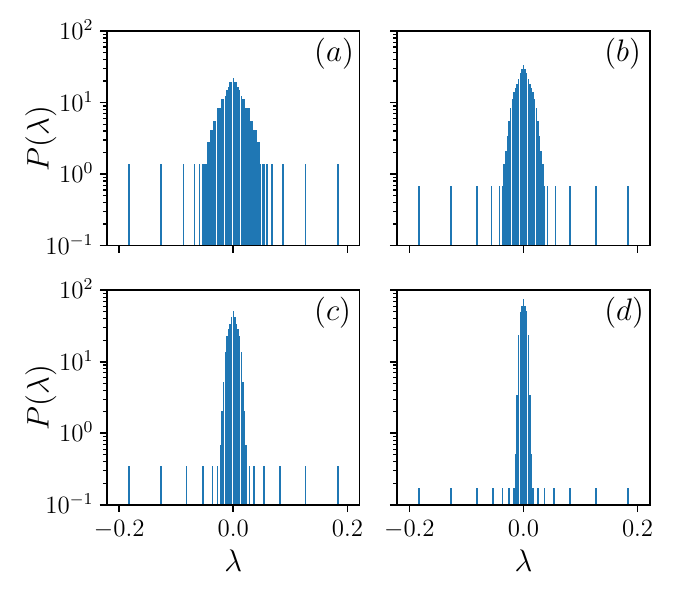}

	\caption {The Lyapunov density $P(\lambda)$ for system size $L = 2^7,2^8,2^9,2^{10} (a,b,c,d)$.  The results are averaged over ${\cal N}=50$ initial states at $\mu = 0$. The integration time is $t_\text{max} = 10^5$.}
	\label{Fig-LS}
\end{figure}

\begin{figure}[t]
\includegraphics[width=1.0\columnwidth]{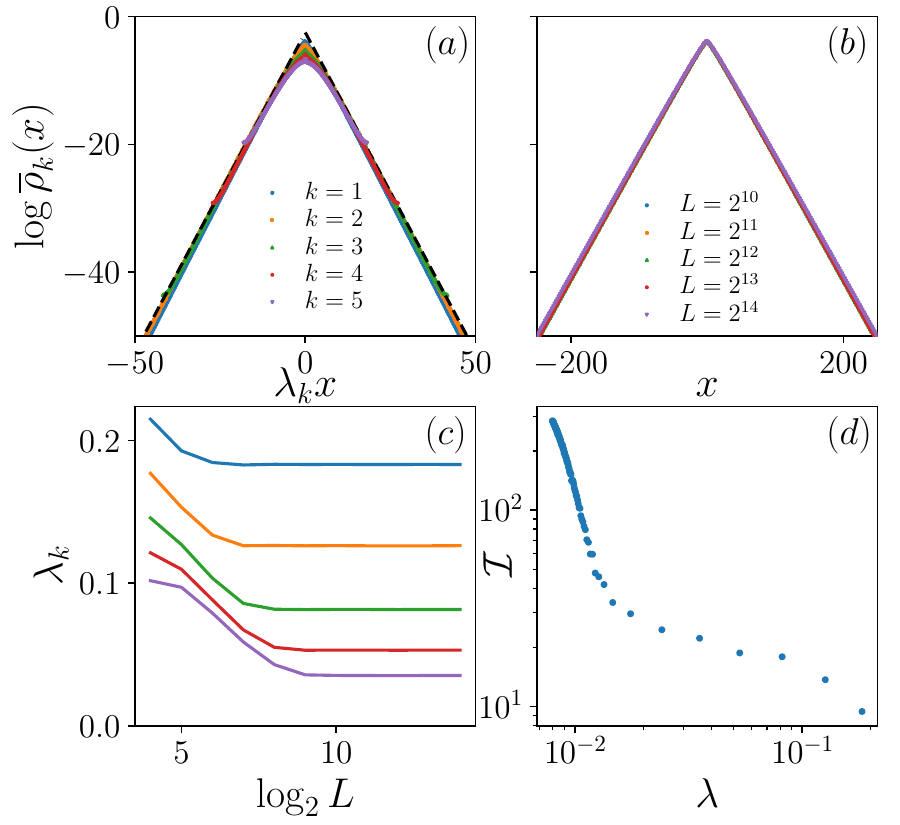}
	\caption{Spatial distribution $\bar{\rho}_k(x)$ of Lyapunov vectors  corresponding to :
    (a) Different leading Lyapunov exponents $\lambda_k$ for $L = 2^{10}$; (b) The largest Lyapunov exponents for different system sizes $L$. (c) The five largest Lyapunov exponents versus $L$. (d) IPR ${\cal I}_k $ of (the first $100$) Lyapunov vectors versus corresponding Lyapunov exponents $\lambda$ for system size $L=2^{10}$. The results are averaged over ${\cal N}=50$ initial states at $\mu = 0$. The black dashed line in (a) indicates $\bar{\rho}_k(x)\sim e^{-\lambda_k |x|}$.
	The integration time $t_\text{max} = 10^6$.}
	\label{Fig-LVector}
\end{figure}

{\it Lyapunov spectrum and Lyapunov vectors.-- }
Using the method of Benettin et al.~\cite{benettin1980lyapunov-Lyp} and facilitating the explicitness and simplicity of dynamical map, we calculate the Lyapunov spectrum $\{\lambda_k\}$ (in descending order) and the corresponding (backward) Lyapunov vectors $\boldsymbol{\psi}_k$. Note that the spatial components $\boldsymbol{\psi}_{k,x}\in \mathbb R^3$ are tangential to unit sphere at the initial point $\boldsymbol{S}_x$.

We simulated very long trajectories up to $t_{\rm max}=10^5$, starting from random initial conditions (at fixed $\mu$) and verified that 
$\lambda_k$ are independent of the initial condition, suggesting that $S$ is the only conserved quantity left upon integrability breaking. To begin, we present Lyapunov spectral density $P(\lambda)=\frac{1}{L}\langle \sum_k \delta(\lambda-\lambda_k)\rangle$ in Fig.~\ref{Fig-LS} where, to improve statistics, the results are averaged $\langle\cdot\rangle$ over ${\cal N}$ initial states sampled from canonical ensemble \eqref{eq-gc} at $\mu = 0$.
For all system sizes studied, the behavior of $P(\lambda)$ exhibit distinct pattern  at different region of $\lambda$: discontinuous regime of converged eigenvalues $\lambda_k|_{L\to\infty}$ at the spectral edge and a smooth distribution at the center. The width of the central spectral peak appears to decrease with increasing $L$. These results suggest a conclusion that $\lim_{L\to\infty}P(\lambda)=\delta(\lambda)$, hence a majority of Lyapunov exponense vanish, while a discrete sequence $\lambda_k$ survives the thermodynamic limit $L\to\infty$. To further corroborate the discontinuous structure and convergence of eigenvalues near the spectral edge,  we plot in Fig.~\ref{Fig-LVector} (c)
the first five Lyapunov exponents $\lambda_{k}$ as a function of $L$. 

Next, we study the structure of the corresponding (backward) Lyapunov vectors $\boldsymbol{\psi}_k$. A quantity of interest is their spatial distribution 
$\rho_k(x) = |\boldsymbol{\psi}^{t_{\rm max}}_{k,x}|^2$.
In the numerical simulations, we take the moving time average 
\be
\log\overline{\rho}_{k}(x)=\frac{1}{\Delta}\sum_{T=t_{\text{max}}-\Delta+1}^{t_{\text{max}}}\log\left\langle \frac{1}{2}\left(|\boldsymbol{\psi}_{k,x}^{T}|^{2}+|\boldsymbol{\psi}_{k,x+1}^{T}|^{2}\right)\right\rangle 
\ee
 over simulation times in interval of width
 $\Delta=1000$, as well as over $\mathcal N$ initial states.
Spatial distribution $\bar{\rho}_k(x)$ of leading Lyapunov vectors are shown in Fig.~\ref{Fig-LVector} (a).
Interestingly, Lyapunov vectors are localized around the local perturbation at $x = 0$.
Furthermore, $\bar{\rho}_k(x)$ for different $k$ approximately overlap when plotted as a function of $\lambda_k x$ for $x \gg 0$, implying that localization lengths are proportional to the inverse of Lyapunov exponents, i.e., 
\be
\label{main}
\bar{\rho}_k(x)\asymp e^{-\lambda_k |x|}\equiv e^{-|x|/\xi_k},\qquad \xi_k = 1/\lambda_k.
\ee
A similar scaling is also observed for different system sizes (Fig.~\ref{Fig-LVector} (b)), for the leading Lyapunov vector $\boldsymbol{\psi}_1$.

Additionally, we also consider the inverse participation ratio (IPR) of $P(x)$, 
${\cal I}_k = \left\{\sum_{x}[\bar{\rho}_k(x)]^2\right\}^{-1}$, assuming Lyapunov vectors to be normalized.
Fig.~\ref{Fig-LVector} (b) shows ${\cal I}_k$ for the first $100$ Lyapunov vectors for $L = 2^{10}$.
IPR is a good empirical indicator of localization length. We observe that ${\cal I}_k$ increase for decreasing $\lambda_k$, which suggests that the Lyapunov vectors corresponding to vanishing exponents become delocalized/extended.

\begin{figure}[t]
	\includegraphics[width=\columnwidth]{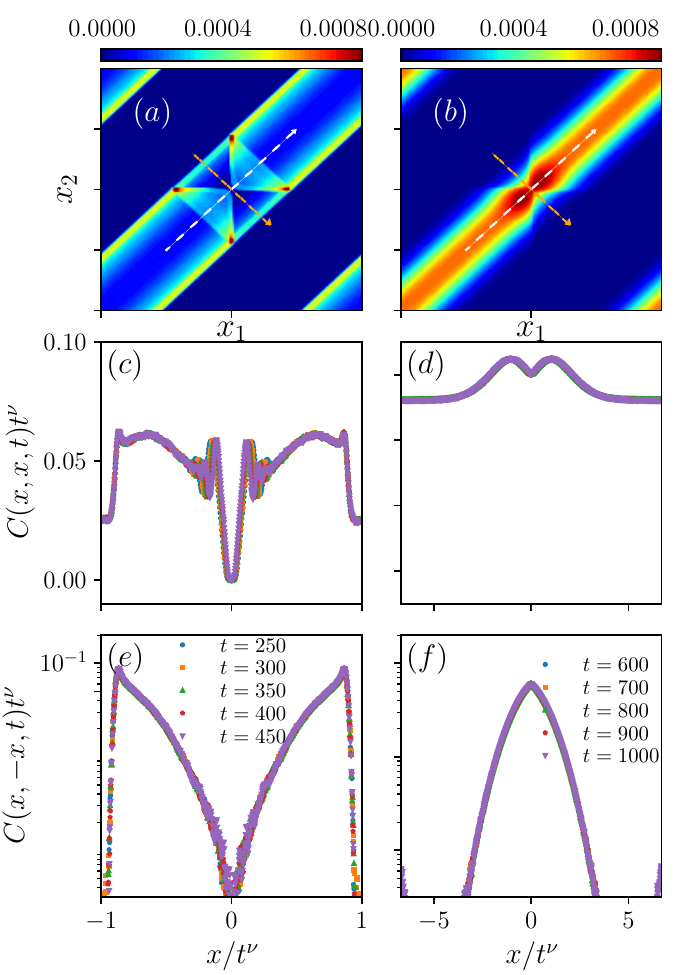}

	\caption{Spin-spin correlation function $C(x_1,x_2,t)$. Heatmaps for $\mu = 0.5,\ t = 250$ (a), and $\mu = 0.0,\ t = 600$ (b). [(c)(e)] : $C(x,x,t)t^\nu$ and $C(x,-x,t)t^\nu$ versus $x/t^\nu$ for $\mu = 0.5$, $\nu = 1$. [(d)(f)]: Similar to [(c)(e)] but for $\mu = 0$, $\nu = 2/3$.
 The results are averaged over $5\times10^7$ initial states for system size $L = 2^{11}$.
	}
	\label{Fig-Ct}
\end{figure}

\begin{figure}[t]
	\includegraphics[width=0.8\columnwidth]{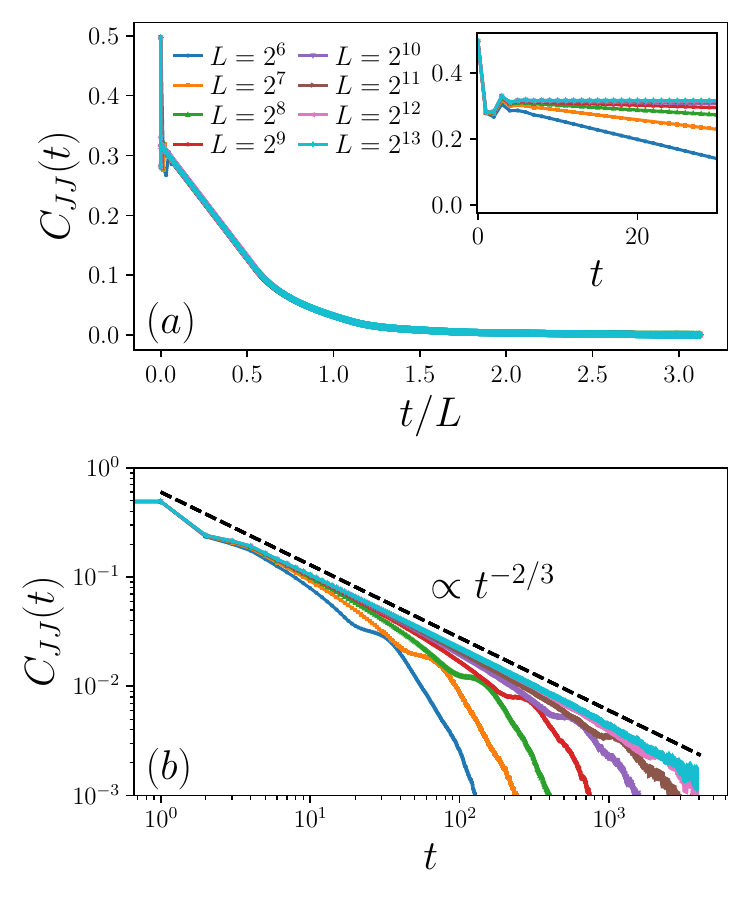}

	\caption{Current-Current correlation function $C_{JJ}(t)$ versus $t/L$ for $\mu = 0.5$ (a), or versus $t$ for $\mu = 0$ (b). Inset of (a): $C_{JJ}(t)$ at short time scale. Results are averaged over ${\cal N}=10^5$ initial states.
	}
	\label{Fig-Jt}
\end{figure}

To qualitatively explain our main finding (\ref{main}) we provide a heuristic argument. Starting from a initial infinitesimal deviation at $x = x^*$, $\delta\boldsymbol{X}^{0}=(0,\cdots,\delta\boldsymbol{S}_{x^*}^{0},\cdots,0)$, we consider the deviation at time $t$. The leading contribution is given by
\be\label{eq-d2t}
d(t)\equiv|\delta\boldsymbol{X}^{t}|\approx e^{\lambda_{1}t}|\boldsymbol{\psi}_{1,x^*}\cdot \delta\boldsymbol{S}^{0}_{x^*}|.
\ee
In the space-time lattice dynamics the Lieb-Robinson or 'sound' speed (denoted by $v$) is finite, e.g., here $v = 1$. As the integrability breaking is localized on a single site $x = 0$, it only affects the evolution of $\delta \boldsymbol{X}^{t}$ at time $t>x/v$, hence it  it might be reasonable to assume 
$d(t)\sim d(0)e^{\lambda_1(t-\frac{|x|}{v})}$ 
for $t > x/v$. 
Comparison with Eq.~\eqref{eq-d2t} yields
$ 
|\boldsymbol{\psi}_{1,x^*}\cdot\delta\boldsymbol{S}_{x^*}^{0}|\lesssim e^{-\frac{\lambda_{1}|x^*|}{v}}$.
This implies that the localization length of $\boldsymbol{\psi}_1$ is bounded as
$
\xi_1\lesssim\frac{v}{\lambda_{1}},
$
and suggests the scaling (\ref{main}) for $k=1$.
The prediction is verified in Fig.~\ref{Fig-LVector}  (a) (and further in Fig.~S1 (a) of~\cite{SM}),  where excellent agreement is observed.


Based on the heuristic discussion above, which can probably be extended to subleading vectors $k>1$, one may conjecture that exponential localization of Lyapunov vectors corresponding to non-vanishing Lyapunov exponents is a universal feature in locally interacting many-body systems with local integrability-breaking perturbations. In the supplemental material~\cite{SM}, we investigate a locally perturbed classical unitary circuit map~\cite{PhysRevLett.128.134102-Flach} and indeed observe compatible results.

{\it Transport behavior.--\ }
To study the impact of local perturbation on transport behavior, we consider the local spin-spin spatio-temporal correlation function
\be
C(x_{1},x_{2},t)=\langle S_{x_{1}}^{t}S_{x_{2}}^{0}\rangle-\langle S_{x_{1}}^{t}\rangle\langle S_{x_{2}}^{0}\rangle,
\ee
as well as the (extensive) current-current correlation function
\begin{equation}
    C_{JJ}(t) = \langle J^t J^0 \rangle,
\end{equation}
where $J^t$ is the mean (extensive) current (see~\cite{SM} for the definition of $J$) and $\langle \bullet \rangle$ indicates canonical ensemble average. In the integrable model (in the absence of perturbation, $\phi=0$), $C(x,x,t)\propto t^{-1/z}$, $C_{JJ}(t) \propto t^{2/z-2}$, with dynamical KPZ exponent $z=3/2$ for $\mu=0$ and ballistic exponent $z=1$ for $\mu\neq 0$~\cite{krajnik2020kardar}.
As a sketch of the general behavior, Fig.~\ref{Fig-Ct} illustrates the spin-spin correlation functions for the locally perturbed model by showing
$C(x_1,x_2,t)$ at specific times and for different magnetizations, $\mu=0$ (a), and $\mu=0.5$ (b). To explore the effect of perturbation on the transport behavior more thoroughly, we further show the results along two specific directions in $x_1-x_2$ plane, cutting through the perturbation, specifically $C(x,x,t)$ in panel [(c)(d)] and $C(x,-x,t)$ in panel [(e)(f)]. 
For both directions, ballistic transport scaling law $\frac{1}{\tau}C(x/\tau,\pm x/\tau,t/\tau)\simeq C(x,\pm x,t)$ (for $x,t\gg 1$ and fixed $\tau$) 
can be clearly observed in [(d)(e)] at non-zero magnetization $\mu = 0.5$. 
Similarly, at zero magnetization $\mu = 0$, results in [(d)(f)] show a clear anomalous scaling $\frac{1}{\tau^{2/3}}C(x/\tau^{2/3},\pm x/\tau^{2/3},t/\tau)\simeq  C(x,\pm x,t)$ (for $x,t\gg 1$ and fixed $\tau$).
These results indicate that a single integrability breaking impurity does not change the overall transport behavior in the thermodynamic limit which is consistent with
the localization of leading Lyapunov vectors. 
Moreover, the propagation of the effect of local
integrability breaking is given by the same dynamical exponent in both cases, so that a 2-dim scaling function should exist
\be
F(\zeta_1,\zeta_2) = 
\lim_{t\to\infty}\lim_{L\to\infty}
t^{-1/z} C(\zeta_1 t^{1/z},\zeta_2 t^{1/z},t).
\label{eq:F}
\ee
While $\lim_{\eta\to\infty}F(\zeta+\eta,\eta)$ should be given by Pr{\"a}hofer-Spohn scaling function~\cite{PS} $f_{\rm PS}(\zeta)$, the potential universality of the 2-dimensional scaling (\ref{eq:F}) should be investigated 
in future. 

Furthermore, the above results are supported by the numerical simulation of current-current correlations in Fig.~\ref{Fig-Jt}. At $\mu = 0.5$,  $C_{JJ}(t)$, for different $L$, overlap as a function of rescaled time $t/L$ at sufficiently large $t$. This suggests  that in the thermodynamic limit $L\rightarrow \infty$, $C_{JJ}(t)$ does not decay to zero, which is the evidence of ballistic transport. On the other hand, for $\mu = 0$, a power-law decay $C_{JJ}(t) \sim t^{-2/3}$ is observed, in consistent with the KPZ scaling found in the unperturbed system~\cite{krajnik2020kardar}.

{\it Conclusion.-- }
We investigated integrability-breaking  and transport behavior in a discrete classical space-time lattice with a local perturbation. We find a simple and potentially universal structure of the Lyapunov spectrum with a majority of Lyapunov exponents vanishing in the thermodynamic limit. The measure-zero sequence of non-zero Lyapunov exponents correspond to 
exponentially localized Lyapunov vectors, 
with localization lengths inversely proportional to Lyapunov exponents.
Moreover, spin-spin and current-current correlation functions of the conserved Noether charges exhibit a simple 2-parameter scaling with the KPZ dynamical exponent in the SU(2) symmetric case (at zero magnetization).
We conjecture our results to be universal for locally broken integrable systems with local interactions. 
A natural next step would be to consider the anisotropic Landau-Lifshitz lattice~\cite{AnisotropicLLL}, where analogous results with modified dynamical exponents are expected.
Another interesting question is to investigate whether, and in what manner, KPZ scaling breaks down as the number of impurities increases.
It is also expected that our results extend to the domain of quantum systems and integrable spin or qubit chains, which could provide a new perspective on the ongoing debate concerning spin transport in the quantum XXZ model with single impurity \cite{PhysRevB.80.125118-XXZ-transport-impurity,PhysRevB.98.235128-XXZ-transport-impurity}.

{\it Acknowledgement.\ }
JW is financially supported by the Deutsche
Forschungsgemeinschaft (DFG), under Grant No. 531128043, as well as under Grant
No.\ 397107022, No.\ 397067869, and No.\ 397082825 within the DFG Research
Unit FOR 2692, under Grant No.\ 355031190. TP acknowledges support by European Research Council (ERC) through Advanced grant QUEST (Grant Agreement No. 101096208), and Slovenian Research and Innovation agency (ARIS) through the Program P1-0402 and Grants N1-0219, N1-0368.
Additionally, we greatly acknowledge computing time on the HPC3 at the University of Osnabr\"{u}ck, granted by the DFG, under Grant No. 456666331.

\bibliographystyle{apsrev4-1}
\bibliography{Ref.bib}

\clearpage
\newpage

 
\setcounter{figure}{0}
\setcounter{equation}{0}
\renewcommand*{\bibnumfmt}[1]{[S#1]}
\renewcommand{\thefigure}{S\arabic{figure}}
\renewcommand{\theequation}{S\arabic{equation}}

\section*{Supplemental material}
In the supplemental material, we provide more details on numerical simulations, additional numerical results, as well as results verification of ergodicity of the model. 
Moreover, we show results for a different model--classical unitary circuit map with local perturbation. 

\section*{Details on numerical simulations}
\subsection{Definition of the spin current}
To introduce the mean (extensive) spin current, it is helpful to start with the local discrete space-time N\" other current, which is defined as
\be
j_{x}^{t}\equiv j(x,t)=\frac{1}{\tau}(q(x,t+1)-q(x,t)),
\ee
where
\be
q(x,t)\equiv S_x^{t}=\boldsymbol{S}_{x}^{t}\cdot\boldsymbol{e_{3}}.
\ee 
The extensive spin current ${\cal J}(t)$ at times $t$ is given by
\be
{\cal J}(t)=\frac{1}{L}\sum_{x=0}^{\frac{L}{2}-1}\left(j(2x,t)+j(2x+1,t+\frac{1}{2})\right),
\ee
where $t+\frac{1}{2}$ correspond to observables obtained after applying just a half-time step map
$\varPsi^{\rm even}_\tau$ after time $t$.

\subsection*{Calculation of Lyapunov exponents and Lyapunov vectors}
In this section we briefly outline the method we employed in calculation of Lyapunov spectrum and the backward Lyapunov vectors.
For convenience of discussions below, we introduce $\boldsymbol{X}^{t}\equiv(\boldsymbol{S}_{0}^{t},\boldsymbol{S}_{1}^{t},\cdots,\boldsymbol{S}_{L-1}^{t}) \in \cal M$ to denote the system state at time $t$.
Considering an initial state $\boldsymbol{X}^{0}$ drawn from canonical ensemble (Eq.~(5) in the main text), and
infinitesimal variations ($\delta\boldsymbol{X}^{0}\equiv(\delta\boldsymbol{S}_{0}^{0},\delta\boldsymbol{S}_{1}^{0},\cdots,\delta\boldsymbol{S}_{L-1}^{0})$), one has at time $t$:
\begin{equation}\label{eq-dR}
\delta\boldsymbol{X}^{t+1}=\boldsymbol{J}(\boldsymbol{X}^{t})\delta\boldsymbol{X}^{t},
\end{equation}
where $\boldsymbol{J}(\boldsymbol{X}^{t})$  is the Jacobian matrix of the map $\Psi^\prime_\tau(\boldsymbol{X}^{t})$. 
In practice, instead of the three-component vector $\boldsymbol{S} = (S^x, S^y, S^z)$, we consider parametrization in terms of a pair of canonical variables $\boldsymbol{S}_c = (S^z, \arctan(S^y/S^z))$, hence we denite the corresponding many-body map as $\varPsi'_\tau$.

To calculate the Lyapunov exponents and corresponding Lyapunov vectors of the system, we consider a non-singular matrix $\boldsymbol{Y} \in \mathbb{R}^{2L\times k}$ ($k$ denotes the number of Lyapunov exponents of interest), the time evolution of which follows Eq. \eqref{eq-dR},
\begin{equation}
\boldsymbol{Y}^{t}=\boldsymbol{J}(\boldsymbol{X}^{t})\cdots\boldsymbol{J}(\boldsymbol{X}^{1})\boldsymbol{J}(\boldsymbol{X}^{0})\boldsymbol{Y}.
\end{equation}
Performing repeated $QR$-decomposition for each step,
\begin{equation}
\boldsymbol{Y}^{t}=Q(t)\cdot R(t)\cdots R(0),
\end{equation}
then the $j$th Lyapunov exponents $\lambda_{k}$ is given by
\begin{equation}
\lambda_{k}=\lim_{t\rightarrow\infty}\sum_{n=0}^{t}\log(R_{jj}(2n)),
\end{equation}
and the (backward) Lyapunov  vectors $\boldsymbol{\psi}_j$ for $\lambda_{k}$ is given by the $j$th column of $Q(t)$ as $t\rightarrow \infty$. Practically, we run the simulation up to maximum integration time $t_\text{max}$ as explained in the main text.

\begin{figure}[t]
\includegraphics[width=1.0\columnwidth]{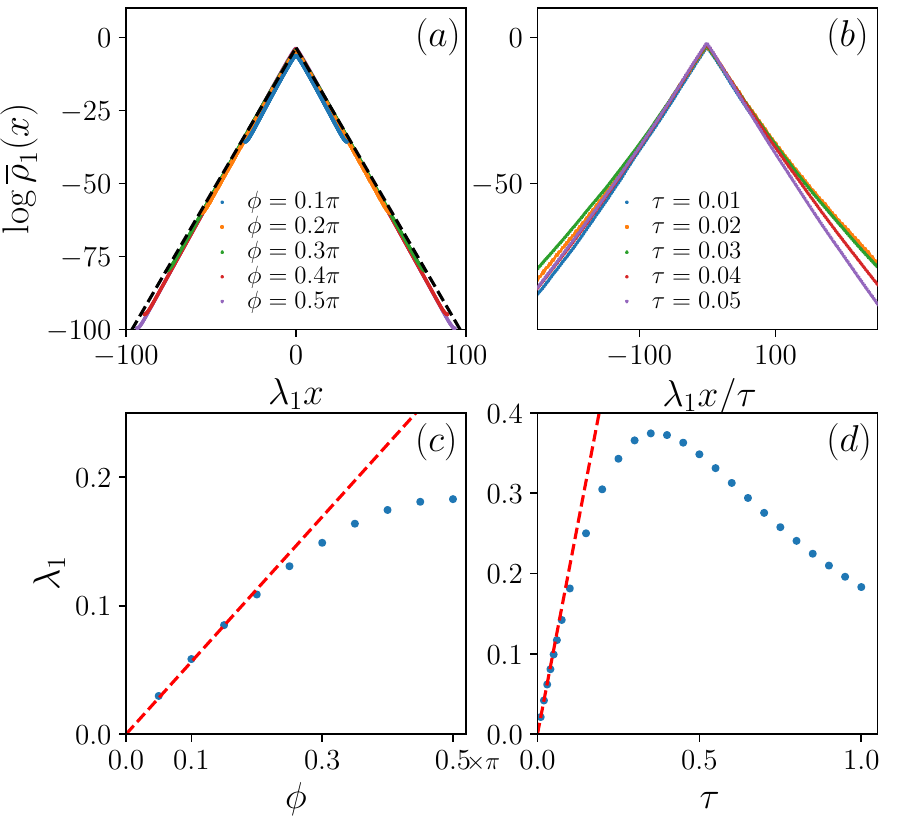}
	\caption{(a) Spatial distribution $\bar{\rho}_1(x)$ of the first Lyapunov vector for different (a) $\phi$ and (b) $\tau$. Largest Lyapunov exponent $\lambda_\mathrm{1}$ versus (c) $\phi$ and (d) $\tau$. The black dashed line in (a) indicates $\bar{\rho}_1(x)\sim e^{-\lambda_1 |x|}$. The red dashed line indicates linear fit as a guidance to the eyes.  The results are averaged over $50$ initial states at $\mu = 0.0$.
    In [(a)(c)] $\tau = 1$ and in [(b)(d)] $\phi = \frac{\pi}{2}$.
   }

	\label{Fig-LVectorT}
\end{figure}

\begin{figure}[t]
	\includegraphics[width=\columnwidth]{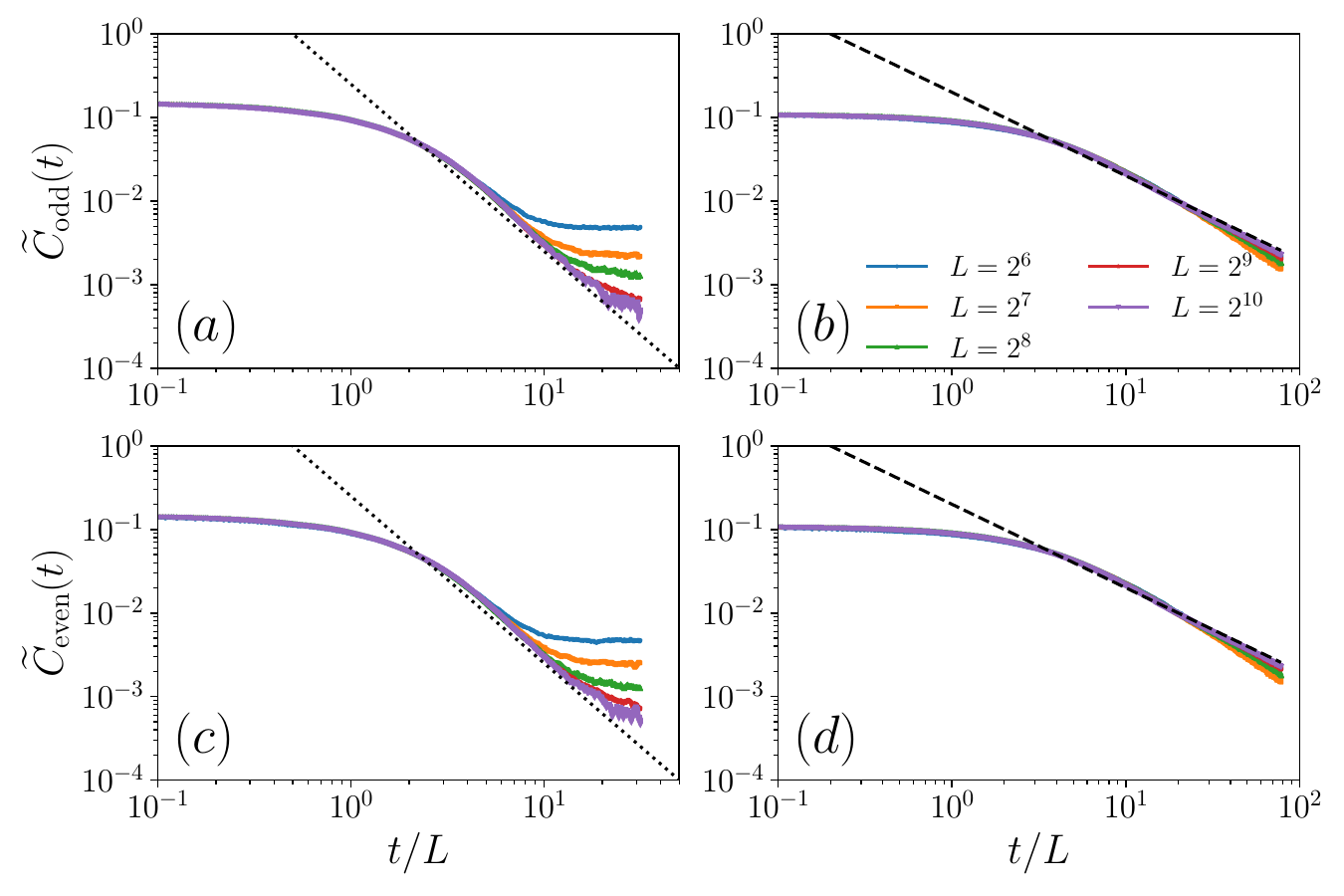}

	\caption{$\widetilde{C}_{\text{odd}}(t)$ and $\widetilde{C}_{\text{even}}(t)$ versus $t/L$ for [(a)(c)] $\mu = 0.0$ and [(b)(d)] $\mu = 0.5$.  The dotted and dashed lines indicate the scaling $\sim t^{-2}$ [(a)(c)] and $\sim t^{-1}$ [(b)(d)], respectively. 
    Results are averaged over ${\cal N}=10^5$ initial states.
	}
	\label{Fig-Erg}
\end{figure}

\section*{Addition numerical results }

In addition to the model parameter $\tau = 1,\ \phi = \frac{\pi}{2}$ considered in the main text, we also explored different parameter values.
Density profiles $\bar{\rho}_1(x)$ of the first Lyapunov vector for different $\phi$ overlap, as a function of $\lambda x$, further supporting our finding $\xi_{k}\propto\frac{1}{\lambda_{k}}$ as in Fig.~2 (a). 

Additionally $\lambda_1$ decreases with decreasing $\phi$, asymptotically as 
$\lambda_1 \propto |\phi| $ for sufficiently small
$|\phi|$. The explanation for this scaling follows straightforwardly from Eq.~(3) in the main text, from which one concludes that the effective perturbation strength $\epsilon_\text{eff}$ decreases as $\phi$ changes from $\frac{\pi}{2}$ to $0$, and that $\epsilon_\text{eff} \propto \phi$ for $\phi \rightarrow 0$.

Similarly, the spatial distribution 
$\bar{\rho}_{1}(x)$ exhibits similar structure [cf. (7) in the main text] for different values of $\tau$ and with a linear dependence of $\lambda_1 \propto \tau$ when $\tau \rightarrow 0$ (Fig.~\ref{Fig-LVectorT}). Note that $\tau\rightarrow 0$ corresponds to the continuous time limit, when the maximal Lyapunov exponents in physical time units can be written as $\lambda_{1}^{0}=\lim_{\tau\rightarrow0}\frac{\lambda_{1}^{\tau}}{\tau}$. The observed linear relation thus implies a positive $\lambda_{1}^{0}$. Similar exponential shape of $\bar{\rho}_1(x)$ at small $\tau$, as shown in Fig.~\ref{Fig-LVectorT} (a) suggests the exponential localization feature persists in the continuous time limit as well.

\section*{Verification of approach to ergodicity under local integrability breaking}

To study the impact of the local integrability breaking term on the ergodicity of the system, we study the time evolution of the following pair of extensive observables
\begin{equation}
Q_{0}^{\text{even}}=\sum_{x=0}^{L/2-1}[q_{0}^{\text{even}}]_{2x},\ Q_{0}^{\text{odd}}=\sum_{x=1}^{L/2}[q_{0}^{\text{odd}}]_{2x-1},
\end{equation}
from some non-equilibrium initial ensemble of initial conditions.  The obserbables $Q_{0}^{\text{even}}$ and $Q_{0}^{\text{odd}}$ are the first two conserved charges of the unperturbed system~\cite{krajnik2020kardar}, 
\begin{gather}
    [q_{0}^{\text{odd}}]_{2x-1}=\log\Big[1+\frac{1}{1+4\tau^{2}}\big(1+2\boldsymbol{S}_{2x+1}\cdot\boldsymbol{S}_{2x}+ \nonumber \\
2\boldsymbol{S}_{2x}\cdot\boldsymbol{S}_{2x-1}+4\tau^{2}\boldsymbol{S}_{2x+1}\cdot\boldsymbol{S}_{2x-1}
\nonumber \\
+2(\boldsymbol{S}_{2x+1}\cdot\boldsymbol{S}_{2x})(\boldsymbol{S}_{2x}\cdot\boldsymbol{S}_{2x-1})\nonumber\\
-4\tau(\boldsymbol{S}_{2x+1},\boldsymbol{S}_{2x},\boldsymbol{S}_{2x-1})\big)\Big ]\ ,
\end{gather}
\begin{gather}
    [q_{0}^{\text{even}}]_{2x}=\log\Big[1+\frac{1}{1+4\tau^{2}}\big(1+2\boldsymbol{S}_{2x+2}\cdot\boldsymbol{S}_{2x+1}+ \nonumber \\
2\boldsymbol{S}_{2x+1}\cdot\boldsymbol{S}_{2x} +4\tau^{2}\boldsymbol{S}_{2x+2}\cdot\boldsymbol{S}_{2x}
\nonumber \\
+2(\boldsymbol{S}_{2x+2}\cdot\boldsymbol{S}_{2x+1})(\boldsymbol{S}_{2x+1}\cdot\boldsymbol{S}_{2x}) \nonumber \\
+4\tau(\boldsymbol{S}_{2x+2},\boldsymbol{S}_{2x+1},\boldsymbol{S}_{2x})\big)\Big ]\ .
\end{gather}

\begin{figure}[t]
\includegraphics[width=1.0\columnwidth]{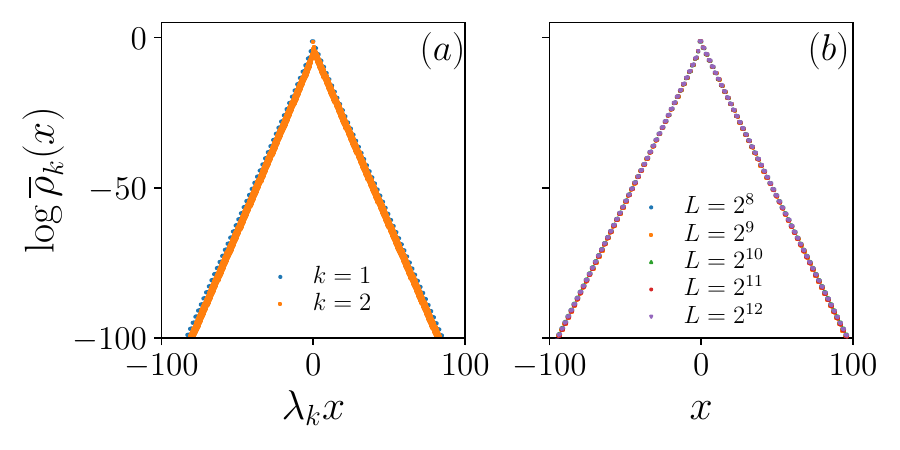}
\includegraphics[width=0.55\columnwidth]{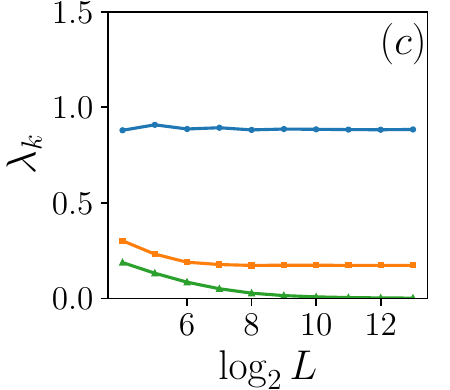}
	\caption{Perturbed unitary circuit map ($n_0 = 1$): spatial distribution $P(x)$ of Lyapunov vectors  corresponding to 
    (a) different Lyapunov exponents $\lambda_k$ (in descending order) for $L = 2^{12}$; (b) largest Lyapunov exponents for different $L$. 
    (c) The first three largest Lyapunov exponents as a function of $L$. The parameter of local nonlinear map  $g = 10,\ \phi = \pi/3$.
    The results are averaged over $96$ initial states and the integration time $t_\text{max} = 10^7$. 
	}
	\label{Fig-LVector-UM1}
\end{figure}

Starting from the non-equilibrium initial ensemble
\begin{equation}\label{ensemble2}
 \rho_{\text{neq}}^{\text{tot}}(\boldsymbol{S}_{1},\cdots,\boldsymbol{S}_{L})=\prod_{x=0}^{L/2-1}\rho_{\mu}(\boldsymbol{S}_{2x})\delta(\boldsymbol{S}_{2x}-\boldsymbol{S}_{2x+1}),
\end{equation}
we study 
\begin{equation}
C_{\text{odd/even}}(t)=\frac{1}{L}\langle Q_{0}^{\text{odd/even}}(t)\rangle^{\prime},
\end{equation}
where the results are averaged $\langle \rangle ^\prime$ over ${\cal N}$ trajectories with initial conditions sampled from the ensemble of Eq.~\eqref{ensemble2}.
In Fig.~\ref{Fig-Erg}, we show the deviation of $C_{\text{odd/even}}(t)$ from its equilibrium value in the canonical ensemble (Eq.~(5) in the main text) $\langle{C}_{\text{odd/even}}\rangle$,
denoted by
\begin{equation}
 \widetilde{C}_{\text{odd/even}}(t)=C_{\text{odd/even}}(t)-\langle{C}_{\text{odd/even}}\rangle.
\end{equation}
For both zero and non-zero magnetization cases, $\widetilde{C}_{\text{odd/even}}(t)$,  for various large system sizes $L$, clearly scale as functions of the rescaled time $t/L$.
This observation is consistent with ergodic behavior at any finite system size, with an ergodic timescale that is proportional with $L$. Furthermore, a power-law scaling is observed, $\widetilde{C}_{\text{odd/even}}(t)\propto t^{-2}$ ($\mu = 0.0$) and 
$\widetilde{C}_{\text{odd/even}}(t)\propto t^{-1}$ ($\mu = 0.5$), the underlying reason for which is left for future investigation.

\begin{figure}[t]
\includegraphics[width=1.0\columnwidth]{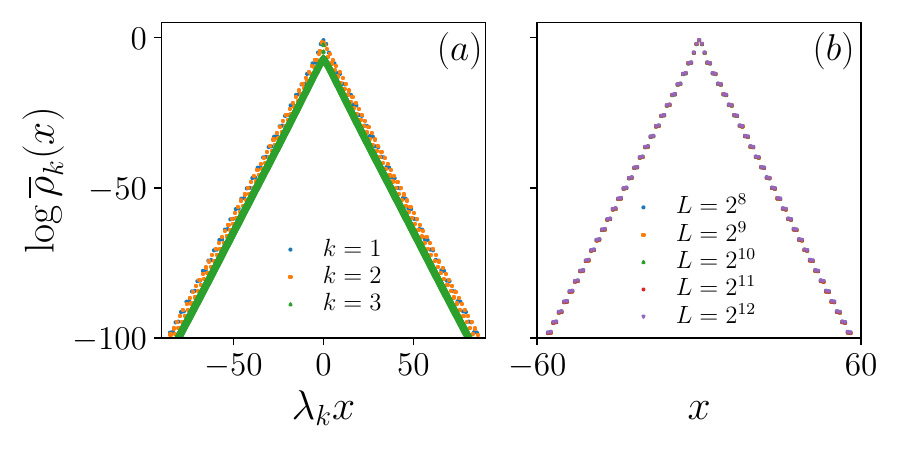}
\includegraphics[width=0.55\columnwidth]{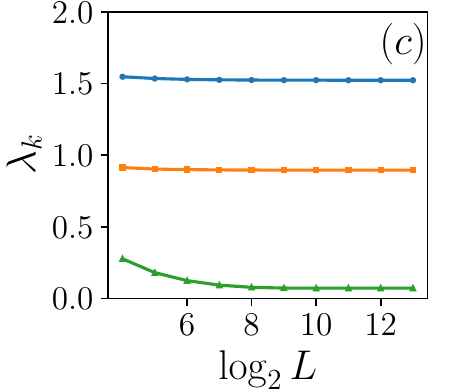}
	\caption{Similar to Fig.~\ref{Fig-LVector-UM1}, but for $n_0 = 2$.
	}
	\label{Fig-LVector-UM2}
\end{figure}

\section*{Results in a classical unitary circuit map with local perturbation}
In addition to the discrete space-time spin lattice studied in the main text, we also consider a classical unitary circuit map.
The map is defined on a 1D lattice of size $L$ with one complex component $\psi_n$ per site $n$. The evolution is performed by subsequent applications of the map
\begin{equation}
\hat{U}_{0}=\prod_{n\in2\mathbb{Z}+1}\hat{C}_{n,n+1}\prod_{n\in2\mathbb{Z}}\hat{C}_{n,n+1}.
\end{equation}
Here
\begin{equation}
    \hat{C}_{n,n+1}\left(\begin{array}{c}
\psi_{n}(t)\\
\psi_{n+1}(t)
\end{array}\right)=\left(\begin{array}{cc}
\cos\phi & \sin\phi\\
-\sin\phi & \cos\phi
\end{array}\right)\left(\begin{array}{c}
\psi_{n}(t)\\
\psi_{n+1}(t)
\end{array}\right).
\end{equation}
The map generated by $\hat{U}_0$ is integrable and symplectic. In fact it is linear and represents a discrete time analogue of coupled harmonic chain.
To break the integrability, we employ a local nonlinear map $\hat{G}_n$,
\begin{equation}
    \hat{G}_{n}\psi_{n}=e^{ig|\psi_{n}|^{2}}\psi_{n}.
\end{equation}
The perturbed map can now be written as
\begin{equation}
\hat{U}=\prod_{n=1}^{n_{0}}\hat{G}_{n}\hat{U}_{0},
\end{equation}
where, differently from Ref. \cite{PhysRevLett.128.134102-Flach}, the local map is only applied to $n_0$ sites. Here we consider $n_0 = 1, 2$.  
We use periodical boundary conditions $\psi_{N+1} = \psi_1$ and the initial conditions for the magnitues of $\psi_n$ are drawn from an exponential distribution and their phases from a uniform distribution in $[0,2\pi]$.
We thus sample initial states from the following `canonical ensemble' $\rho^{\rm tot}(\psi_1,\ldots,\psi_N) = \prod_{n=1}^n e^{-|\psi_n|}$.
The state vector is further rescaled such that norm density $\frac{1}{N}\sum|\psi_{n}|^{2}=1$.

In Figs.~\ref{Fig-LVector-UM1} and \ref{Fig-LVector-UM2} , we show spatial distribution of first several Lyapunov vectors and their corresponding Lyapunov values.
Similar to the results in main text, Lyapunov vectors are also localized around the local perturbation with their localization length $\xi_k \propto \frac{1}{\lambda_k}$. The localization persists as system size $L$ increases.
Additionally, the first few Lyapunov exponents $\lambda_k$ appear to saturate for sufficiently large $L$. In contrast to the lattice Laundau-Lifshitz model of the main text, here it seems we only have a finite number (two for $n_0=1$ and three for $n_0=2$) of non-zero Lyapunov exponents (for any sufficiently large $N$), which may be a consequence of linearity of unperturbed model.

The results in the perturbed unitary circuit map further support the universality of our main finding in local perturbed many-body maps.


\end{document}